\documentclass[conference,a4paper]{IEEEtran}
\usepackage{cite}
\usepackage{amsmath,amssymb,amsfonts}
\usepackage{graphicx}
\graphicspath{{./figs/}}
\usepackage{url}
\usepackage{xcolor}
\usepackage{hyperref}
\hypersetup{hidelinks}

\IEEEoverridecommandlockouts

\begin{document}

% The paper's citation numbering follows the verified camera-ready bibliography.
\nocite{wangzhang2026full,yin2018,erc8004-forum,eip1,a2a,google-donation,a2a-governance,gitvote,reineke2025,dahlander2011,tanaka2015,cennamo2025,allen2025treasury,maharajan2026,unesco2021recommendation,unescoEIA,unescoRAM,women4ethicalai,hannigan2019topic,guidi2025bitcoin}

\title{Revisiting Sustainability by Design in AI Protocol Governance: An Empirical Review of Comparative Open and Corporate-Led Standards for the SDGs%
\thanks{An earlier version of this paper was presented at the 8th International Conference on Management and Organization (ICMO 2026), held in Pr\v{z}no, Montenegro, from 3 to 6 June 2026. The conference, themed ``Sustainability by Design -- Rethinking Strategy, People \& Digital Futures,'' was hosted by the Montenegrin Academy of Management (MAOM), in collaboration with the Slovenian Academy of Management, the Croatian Academy of Management, and the Faculty of Economics Podgorica at the University of Montenegro. We are grateful to the conference organizers and participants for their valuable comments and constructive feedback, which helped improve this paper. DOI: \url{https://doi.org/10.70314/is.2026.ai4sdg.1}.}}

\author{
\IEEEauthorblockN{Yutian Wang}
\IEEEauthorblockA{
\textit{Duke Kunshan University}\\
Suzhou, China
}
\and
\IEEEauthorblockN{Luyao Zhang\thanks{Corresponding author: Luyao Zhang (lz183@duke.edu), Digital Innovation Research Center and Social Science Division, Duke Kunshan University. Address: Duke Avenue No.8, Kunshan, Suzhou, Jiangsu, China, 215316.}}
\IEEEauthorblockA{
\textit{Duke Kunshan University}\\
Suzhou, China}

}

\maketitle

\begin{abstract}
As artificial intelligence (AI) agents enter production infrastructure, the protocols that let them interoperate also shape how that infrastructure is governed and sustained. This paper revisits evidence from our earlier comparative governance study through a Sustainability by Design (SbD) lens. That empirical study compares two standards for AI-agent interoperability: Ethereum Request for Comments 8004 (ERC-8004), developed through an open Ethereum standards process, and Google's Agent2Agent (A2A), governed by a corporate consortium. It uses a large language model (LLM)-powered pipeline that combines automated annotation, neural topic modeling, and multi-layer network analysis to examine socio-technical power structures across large public governance repositories. The study reports two contrasting governance architectures. ERC-8004 uses open participation, public review, and decoupled deployment, whereas A2A assigns binding authority to an eight-seat Technology Steering Committee. It also finds distinct innovation infrastructures. The ERC community concentrates on constitutive questions about trust and security, including what to build and why, while the consortium spreads attention across executive engineering questions about how to implement, document, and deliver the protocol. Both communities show high participation inequality, and corporate contributors span roughly twice as many themes as ERC contributors. This review asks: (1) How do these governance architectures produce distinct SDG-relevant signatures? (2) What design principles do those signatures suggest for sustainable AI governance? We interpret the reported institutional, discursive, and network patterns in relation to SDGs 8, 9, 10, 11, 12, 16, and 17. The analysis identifies capacities for transparency, participation, contestability, and cross-protocol coordination. It does not treat those capacities as evidence of achieved SDG impact. We argue that sustainable AI infrastructure needs a corrective feedback loop between designed charters and governance in practice, an institutional commitment associated with SDG 16 (Peace, Justice, and Strong Institutions). The review brings together the prior study's computational evidence, organizational research, and sustainable development to derive design principles for sustainable AI governance intended for scholars, policymakers, and protocol designers.
\end{abstract}


%================================================================
\section{Introduction}

Governance for the Sustainable Development Goals (SDGs) includes the institutions behind AI infrastructure. When autonomous agents negotiate tasks across organizational boundaries, interoperability protocols distribute authority, innovation work, and participation. Research on AI for the SDGs more often evaluates applications than the protocols beneath them.

\begin{figure*}[t]
  \centering
  \includegraphics[width=0.84\textwidth]{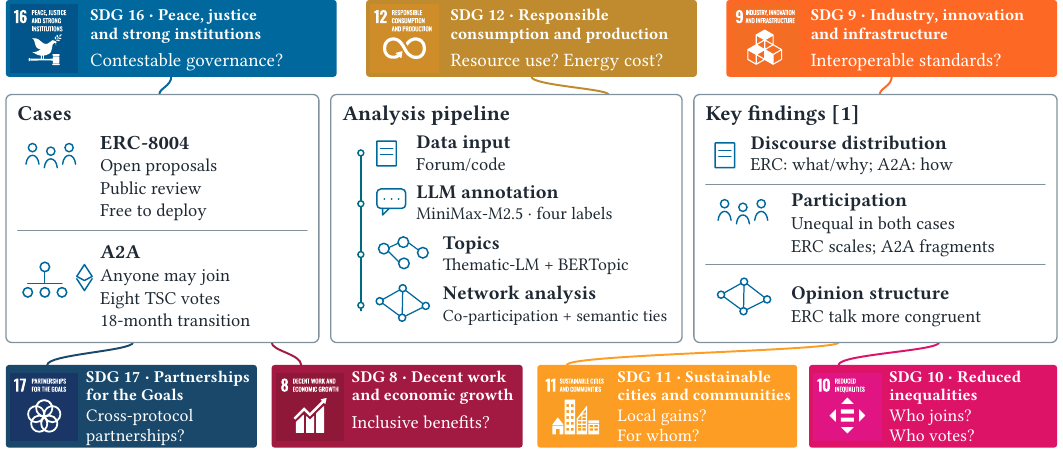}
  \caption{White panels show the earlier comparison~\cite{wangzhang2026full}, with governance details checked against~\cite{erc8004-forum,eip1,a2a-governance,gitvote}. Color panels pose SDG questions, not measured outcomes or protocol energy footprints. SDG icons: United Nations (\url{https://sdgs.un.org/goals}); use does not imply UN endorsement.}
  \label{fig:opening}
\end{figure*}

We address this gap by reviewing our earlier comparative governance study~\cite{wangzhang2026full} through a \textit{Sustainability by Design} (SbD) lens. In this review, an \emph{SDG governance signature} is an observable institutional, discursive, or participation pattern that bears on an SDG-relevant capacity. A signature does not by itself demonstrate SDG achievement. Figure~\ref{fig:opening} summarizes the paper's central analytical claim: governance design sets rules for access, authority, and transparency, and public records then show what communities discuss, who participates, and how participants connect. We use these signatures to consider implications for SDGs~8, 9, 10, 11, 12, 16, and 17.

The earlier study compares two protocols that address the same AI-agent interoperability problem and have public governance records~\cite{yin2018}. The data and analysis materials are available at \url{https://github.com/Global-Nomad-Nexus/erc8004-a2a-case-study}. Ethereum Request for Comments 8004 (ERC-8004), developed through an open Ethereum standards process, is an open Ethereum standard that enables agents to discover, choose, and interact across organizational boundaries without pre-existing trust~\cite{erc8004-forum}.

Its proposal and review process is public, while deployment is separate from standards discussion~\cite{erc8004-forum,eip1}. Google's Agent2Agent (A2A) is an open protocol for communication and interoperability between opaque agentic applications. Google launched it in April 2025 and donated it to the Linux Foundation in June 2025~\cite{a2a,google-donation}. Its charter permits anyone to contribute or become a maintainer, while an eight-seat Technology Steering Committee (TSC) holds initial voting authority. The 18-month provision concerns future TSC composition~\cite{a2a-governance,gitvote}. This design holds the technical domain broadly constant while comparing open entry in both cases with different routes to binding decisions~\cite{a2a-governance,reineke2025}.

The review addresses two questions:
\begin{itemize}
  \setlength{\itemsep}{0pt}
  \setlength{\parskip}{0pt}
  \setlength{\topsep}{2pt}
  \item \textbf{RQ1:} How do contrasting AI protocol governance architectures produce distinct SDG-relevant signatures?
  \item \textbf{RQ2:} What design principles do these signatures imply for sustainable AI governance?
\end{itemize}

We reinterpret this prior empirical comparison through SbD, develop a signature framework covering seven goals, and propose design principles for sustainable AI governance.

%================================================================
\section{Reviewed evidence and analytical lens}

The earlier empirical study reports three types of evidence~\cite{wangzhang2026full}: decision architectures that document entry and binding authority, discourse distributions that distinguish constitutive design from implementation, and participation networks that measure concentration and cross-standard coordination. The original study contains the methodological details, validation procedures, and robustness checks.

This review does not repeat that empirical analysis. It interprets the reported evidence as SDG-relevant governance capacities, considers the design corrections those capacities may require, and keeps institutional capacity distinct from achieved societal impact.

%================================================================
\section{Findings}

\subsection{Governance Architecture Signatures}

The prior study reports contrasting governance architectures. ERC-8004 has a public proposal and review process and separates deployment from standards discussion. A2A reserves binding authority for its eight-seat TSC~\cite{a2a-governance}. EIP-1's rough-consensus provision concerns Core EIPs~\cite{eip1}; A2A's TSC makes binding decisions under its charter and GitVote procedure~\cite{a2a-governance,gitvote}.

\begin{figure*}[htbp]
  \centering
  \includegraphics[width=0.86\textwidth]{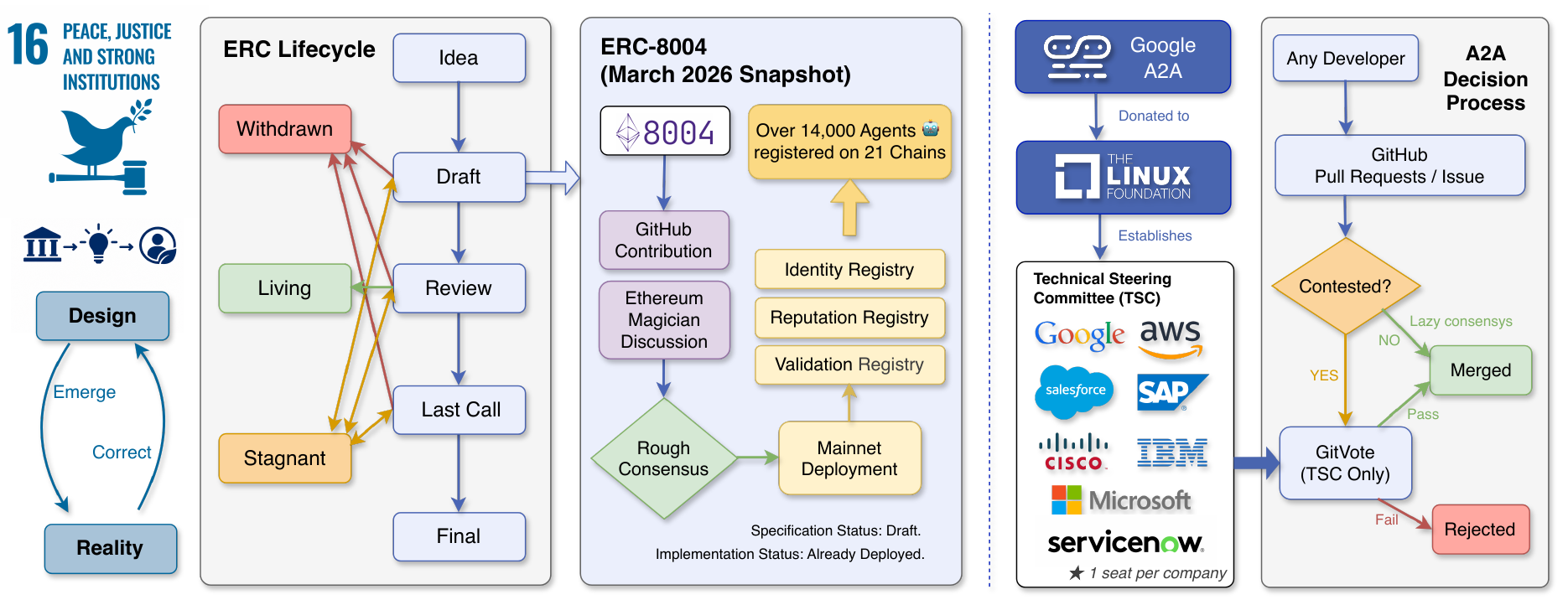}
  \caption{The prior study's contrasting decision architectures~\cite{wangzhang2026full}, with this review's SDG~16 Design~$\rightarrow$~Emerge~$\rightarrow$~Reality~$\rightarrow$~Correct loop.}
  \label{fig:decision}
\end{figure*}

Figure~\ref{fig:decision} adds this review's interpretation. Its left annotation presents a \emph{Design~$\rightarrow$~Emerge~$\rightarrow$~Reality~$\rightarrow$~Correct} loop. The loop compares authority in practice with the stated rules, then uses that comparison to revise the governance design.

\subsection{Different Objects of Deliberation}

The prior study finds that A2A devotes nearly twice the share of its discourse to Process arguments as ERC-8004 ($25.4\%$ vs.\ $13.9\%$; $p < .001$). Within ERC-8004, Technical arguments dominate the design stage, while Process arguments rise during ratification ($p < .001$). ERC-8004 participants focus on what should define the protocol, especially trust mechanisms, security, identity, and specification. A2A participants spread their attention across implementing, documenting, integrating, and shipping an established protocol.

The prior study's thematic analysis shows that ERC-8004 centers on Trust \& Security and Protocol Specification, whereas A2A covers Documentation, software development kits (SDKs), Tooling, and other execution themes. Both communities have high participation inequality. A small core accounts for most interactions, while nearly half of A2A participants and one third of ERC-8004 participants remain isolated. Corporate contributors span roughly twice as many themes as ERC contributors, consistent with specialization by design in A2A and specialization by attrition in ERC-8004~\cite{dahlander2011}.

\begin{figure}[t]
  \centering
  \includegraphics[width=0.88\textwidth,height=0.26\textheight,keepaspectratio]{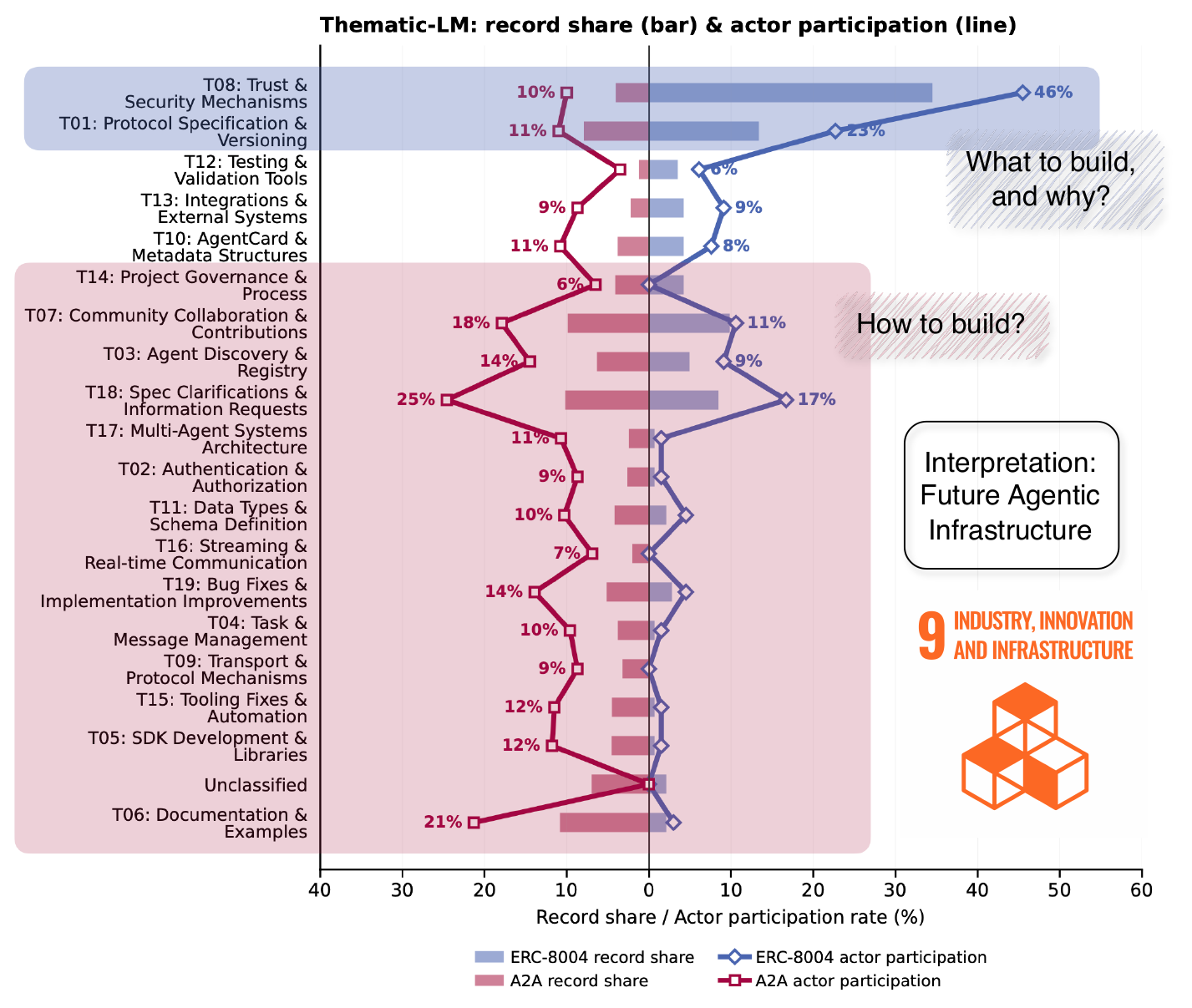}
  \caption{Thematic-LM record share (bars) and actor participation (lines) for what/why and how themes~\cite{wangzhang2026full}. The SDG~9 marker is this review's interpretation, not measured impact.}
  \label{fig:thematic}
\end{figure}

This review interprets that deliberative split in relation to SDG~9 (Industry, Innovation and Infrastructure). Figure~\ref{fig:thematic} presents the thematic differences reported in the prior study~\cite{wangzhang2026full} and keeps them separate from this review's SDG association.

\subsection{Connectivity Depends on the Scale of Observation}

The prior study finds different structures at the protocol and ecosystem scales. In the comparison between ERC-8004 alone and A2A, both public networks are fragmented and participation is concentrated. When the ERC corpus expands from 142 ERC-8004 records to 1,664 records across 34 related AI protocols, repeat contributors and cross-standard discussions connect previously separate threads. The ERC giant-component ratio rises from 0.328 to 0.917. A2A's public trace remains fragmented at 0.285 in the expanded analysis~\cite{wangzhang2026full}.

In this review, the difference between scales shows that protocol-level fragmentation can coexist with ecosystem-level coordination. It also means that connectivity cannot be used as a proxy for inclusion.

%================================================================
\section{Discussion}

The earlier study identifies different governance capacities and risks. ERC-8004 makes constitutive deliberation visible and allows anyone to participate, while A2A coordinates implementation through formal corporate authority. The comparison does not show that open standards advance sustainability or that a corporation obstructs it. We use the SDGs as an assessment lens. They are not evidence that either protocol has achieved societal impact.

\subsection{Extended SDG and Policy Implications}

Our interpretation extends the institutional comparison beyond SDG~16. ERC-8004 makes questions of trust, security, and identity open to public contestation, while A2A's TSC organizes work across documentation, tooling, and integration. In relation to SDG~9, these patterns suggest two requirements. Open protocols need reliable procedures for resolution and maintenance, and corporate-led protocols need auditable routes for outside challenge. The high isolation reported in both cases is also relevant to SDGs~8, 10, and 11, because access, governance labor, and connection to decisions are distinct outcomes.

The 34-protocol comparison shows why our interpretation must account for scale. Cross-standard contributors form an ecosystem-level capacity for partnership that is relevant to SDG~17, but concentrated participation persists. Connectivity and inclusion therefore require separate measures. Public records offer an analogy to SDG~12 traceability, not proof of sustainable consumption. Environmental footprints need separate measurement: 34 large U.S. Bitcoin mines consumed an estimated 32.3~TWh during August 2022--July 2023~\cite{guidi2025bitcoin}. That proof-of-work result cannot be assigned to ERC-8004 or A2A.

The United Nations Convention on the Law of the Sea (UNCLOS) provides a boundary case for coordination. The law of the sea coordinates states in a cross-border domain through durable institutions and negotiated obligations~\cite{tanaka2015}. AI protocols, by contrast, can fork, recombine, and migrate without territorial continuity. Sustainable AI governance can draw on the durability of international coordination without adopting a state-centered form. Platform regulation~\cite{cennamo2025} and blockchain treasury governance~\cite{allen2025treasury} frame authority and accountability; a UNICEF pilot illustrates one funding route for open digital public goods~\cite{maharajan2026}.

\subsection{UNESCO AI Ethics and Governance Frameworks}
\vspace{-0.5\baselineskip}

UNESCO's 2021 Recommendation on the Ethics of Artificial Intelligence, adopted by 193 Member States, calls for inclusive multi-stakeholder governance, responsibility, auditability, traceability, transparency, non-discrimination, and human oversight~\cite{unesco2021recommendation}. We use these principles to interpret the empirical comparison. ERC-8004's permissionless entry and public deliberation records correspond more closely to participation and auditability. A2A's initial eight corporate TSC voting seats concentrate formal authority despite open contributor and maintainer pathways, creating tension with inclusiveness even as the TSC coordinates implementation. The participation results also show that open entry does not ensure inclusion: both cases remain concentrated.

UNESCO's Ethical Impact Assessment (EIA) evaluates individual AI systems across their life cycle, whereas its Readiness Assessment Methodology (RAM) evaluates institutional capacity at the country level~\cite{unescoEIA,unescoRAM}. Neither tool directly assesses protocol communities. We draw an analogy with EIA and RAM; we do not use either instrument to assign scores. Publicly traceable forum and GitHub records allow external review, contestation, and institutional diagnosis. ERC-8004 provides a broader observable record. A2A's public repository is also observable, but outsiders have less access to binding authority. Under this interpretation, sustainable governance related to SDG~16 requires records that can be assessed and contested, as well as a designed charter.

The \emph{Design~$\rightarrow$~Emerge~$\rightarrow$~Reality~$\rightarrow$~Correct} loop in Figure~\ref{fig:decision} is a protocol-governance form of UNESCO's monitoring logic. The \emph{Reality~$\rightarrow$~Correct} step connects impact monitoring and external review to institutional revision. ERC-8004's public discussion provides a feedback channel, but the protocol still needs explicit procedures for resolution and maintenance. A2A has clearer executive authority, but it needs an auditable route for outsiders to contest and revise binding rules.

Participation inequality further shows the difference between formal access and inclusion in practice. UNESCO's Women4EthicalAI platform seeks equal representation in AI design, use, and deployment~\cite{women4ethicalai}. ERC-8004 removes a formal barrier to entry, while A2A concentrates initial TSC voting authority among corporate members. Neither finding demonstrates gender equality because the dataset does not identify gender. Protocol stewards should report representation, isolation, coordination work, and access to binding decisions as separate measures.

At the ecosystem scale, cross-protocol ties create governance capacity relevant to SDGs~9 and 17. Because these ties coexist with concentrated participation, connectivity and inclusion should remain separate governance indicators.

These conclusions are limited by the use of public records, the focus on two governance forms, dependence on natural language processing models~\cite{hannigan2019topic}, and culturally contingent ideas of fairness. Further research should examine other protocol families and whether corrective mechanisms improve participation and decision quality. Sustainable AI governance needs technical openness and institutions that keep authority observable, connect deliberation across protocol and ecosystem scales, and correct inequalities that those institutions themselves produce.

%================================================================
\section*{Acknowledgments}
L.Z. acknowledges the support from the National Natural Science Foundation of China (NSFC) under the project titled ``Trust Mechanism Design on Blockchain: An Interdisciplinary Approach of Game Theory, Reinforcement Learning, and Human-AI Interactions'' (Grant No. 12201266). Y.W. gratefully acknowledges the Summer Research Scholar Program and the Office of Academic Services at Duke Kunshan University for support.

\bibliographystyle{IEEEtran}
\bibliography{refs}

@article{reineke2025,
  author  = {Reineke, Philipp and Katila, Riitta and Eisenhardt, Kathleen M.},
  title   = {Decentralization in Organizations: {A} Revolution or a Mirage?},
  journal = {Academy of Management Annals},
  volume  = {19},
  number  = {1},
  pages   = {298--342},
  year    = {2025}
}

@misc{eip1,
  author       = {Becze, Martin and Jameson, Hudson and others},
  title        = {{EIP-1}: {EIP} Purpose and Guidelines},
  year         = {2015},
  url          = {https://eips.ethereum.org/EIPS/eip-1}
}

@misc{erc8004-forum,
  author       = {Crapis, Davide},
  title        = {{ERC-8004}: Trustless Agents},
  howpublished = {Ethereum Magicians Forum, topic 25098},
  year         = {2025},
  url          = {https://ethereum-magicians.org/t/erc-8004-trustless-agents/25098}
}

@misc{gitvote,
  author       = {{a2aproject}},
  title        = {{.gitvote.yml}},
  howpublished = {GitHub},
  year         = {2025},
  url          = {https://github.com/a2aproject/A2A/blob/main/.gitvote.yml}
}

@misc{a2a,
  author       = {{A2A Project}},
  title        = {{Agent2Agent (A2A) Protocol}},
  year         = {2025},
  url          = {https://github.com/a2aproject/A2A}
}

@misc{a2a-governance,
  author       = {{A2A Project}},
  title        = {{GOVERNANCE.md}},
  year         = {2025},
  url          = {https://github.com/a2aproject/A2A/blob/main/GOVERNANCE.md}
}

@misc{google-donation,
  author       = {{Google Cloud}},
  title        = {Google {Cloud} Donates {A2A} to {Linux Foundation}},
  year         = {2025},
  url          = {https://developers.googleblog.com/en/google-cloud-donates-a2a-to-linux-foundation/}
}

@book{yin2018,
  author    = {Yin, Robert K.},
  title     = {Case Study Research and Applications: {Design} and Methods},
  edition   = {6th},
  publisher = {SAGE},
  year      = {2018}
}

@misc{wangzhang2026full,
title={Agentic Analysis for Agentic Infrastructure: An {LLM}-Powered Pipeline for Comparative Governance of {DAO} and Corporate {AI} Protocols},
author={Yutian Wang and Luyao Zhang},
year={2026},
eprint={2606.26203},
archivePrefix={arXiv},
primaryClass={cs.AI},
url={https://arxiv.org/abs/2606.26203},
}

@article{cennamo2025,
  title={Platform Regulation: Beyond Power and Size},
  author={Cennamo, Carmelo and Gianiodis, Peter and Kretschmer, Tobias and McIntyre, David and Ozcan, Pinar and Van Alstyne, Marshall},
  journal={Academy of Management Perspectives},
  year={2025},
  doi={10.5465/amp.2025.0261},
  note={doi:10.5465/amp.2025.0261},
  publisher={Academy of Management}
}

@article{hannigan2019topic,
  author  = {Hannigan, Timothy R. and Haans, Richard F. J. and Vakili, Keyvan and Tchalian, Hovig and Glaser, Vern L. and Wang, Milo Shaoqing and Kaplan, Sarah and Jennings, P. Devereaux},
  title   = {Topic Modeling in Management Research: {Rendering} New Theory from Textual Data},
  journal = {Academy of Management Annals},
  volume  = {13},
  number  = {2},
  pages   = {586--632},
  year    = {2019}
}

@article{dahlander2011,
  author  = {Dahlander, Linus and O'Mahony, Siobh{\'a}n},
  title   = {Progressing to the Center: {Coordinating} Project Work},
  journal = {Organization Science},
  volume  = {22},
  number  = {4},
  pages   = {961--979},
  year    = {2011}
}

@book{tanaka2015,
  author    = {Tanaka, Yoshifumi},
  title     = {The International Law of the Sea},
  edition   = {2nd},
  publisher = {Cambridge University Press},
  address   = {Cambridge},
  year      = {2015}
}

@misc{unesco2021recommendation,
  author       = {{UNESCO}},
  title        = {Recommendation on the Ethics of Artificial Intelligence},
  year         = {2021},
  howpublished = {UNESCO General Conference, 41st session},
  url          = {https://unesdoc.unesco.org/ark:/48223/pf0000381137},
  note         = {Accessed: July 19, 2026}
}

@misc{unescoEIA,
  author       = {{UNESCO}},
  title        = {Ethical Impact Assessment},
  year         = {n.d.},
  howpublished = {Global AI Ethics and Governance Observatory},
  url          = {https://www.unesco.org/ethics-ai/en/eia},
  note         = {Accessed: July 19, 2026}
}

@misc{unescoRAM,
  author       = {{UNESCO}},
  title        = {Readiness Assessment Methodology: A Tool of the Recommendation on the Ethics of Artificial Intelligence},
  year         = {2023},
  url          = {https://unesdoc.unesco.org/ark:/48223/pf0000385198},
  note         = {Accessed: July 19, 2026}
}

@misc{women4ethicalai,
  author       = {{UNESCO}},
  title        = {{Women4Ethical AI}},
  year         = {n.d.},
  url          = {https://www.unesco.org/en/artificial-intelligence/women4ethical-ai},
  note         = {Accessed: July 19, 2026}
}

@article{allen2025treasury,
  author  = {Allen, Darcy W. E. and Berg, Chris and Lane, Aaron M.},
  title   = {The Blockchain Treasury Governance Dilemma},
  journal = {Regulation \& Governance},
  volume  = {19},
  number  = {3},
  pages   = {761--776},
  year    = {2025},
  doi     = {10.1111/rego.12659},
  note    = {doi:10.1111/rego.12659}
}

@misc{maharajan2026,
  author       = {Maharajan, Arun},
  title        = {Exploring Sustainable Funding for Digital Public Goods},
  howpublished = {UNICEF Office of Innovation},
  year         = {2026},
  month        = jan,
  day          = {7},
  url          = {https://www.unicef.org/innovation/stories/exploring-sustainable-funding-digital-public-goods}
}

@article{guidi2025bitcoin,
  author  = {Guidi, G. and Dominici, F. and Steinsultz, N. and others},
  title   = {{The} Environmental Burden of the {United States'} Bitcoin Mining Boom},
  journal = {Nature Communications},
  volume  = {16},
  year    = {2025},
  doi     = {10.1038/s41467-025-58287-3},
  note    = {art. 2970, doi:10.1038/s41467-025-58287-3}
}

\end{document}